\documentclass[%
aps
 amsmath,amssymb,
 reprint,%
]{revtex4-1}
\usepackage{accents}
\usepackage{subfigure}
\usepackage{graphicx}
\usepackage{dcolumn}
\usepackage{bm}
\usepackage{appendix}
\begin{document}


\title{A new approach to stochastic acceleration of electrons in colliding laser fields }

\author{Yanzeng Zhang}
\author{Sergei Krasheninnikov}%
\affiliation{%
 Mechanical and Aerospace Engineering Department, University of California San Diego, La Jolla, CA 92093, USA
}%


\begin{abstract}
The mechanism of stochastic electron acceleration in colliding laser waves is investigated by employing proper canonical variables and effective time, such that the new Hamiltonian becomes time independent when the perturbative (weaker) laser wave is absent. The performed analytical analysis clearly reveals the physical picture of stochastic electron dynamics. It shows that when the amplitude of the perturbative laser field exceeds some critical value, stochastic electron acceleration occurs within some electron energy range. The conditions, at which the maximum electron energy gained under stochastic acceleration is greatly exceeding the ponderomotive energy scaling based on the amplitude of the dominant laser, are derived.

\end{abstract}

\maketitle

The electron dynamics in counter-propagating laser waves \cite{bauer1995relativistic,shvets1998superradiant,esarey1997electron,*schroeder1999generation} has attracted a great deal of interest, and the stochastic acceleration was thought to be the reason for high energized electron tails observed in laser-plasma interaction \cite{zaslavskii1968stochastic,rechester1979stochastic,mendonca1982stochasticity,*mendonca1983threshold,rax1992compton,sheng2002stochastic,*sheng2004efficient,bochkarev2018stochastic,forslund1985two}. However, due to the multidimensional spatio-temporal characteristics of the laser waves and strong nonlinearity of the dynamics of relativistic electrons in these waves, the analytic investigations of stochastic electron acceleration in the colliding laser waves in earlier studies have been limited to either non-relativistic case \cite{zaslavskii1968stochastic,rechester1979stochastic} or the stochastic instability near the separatrices using quite complicated multidimensional Hamiltonian approach  \cite{LDLandaubook,mendonca1982stochasticity,*mendonca1983threshold,rax1992compton}. Although the numerical simulations \cite{sheng2002stochastic,*sheng2004efficient,bochkarev2018stochastic} shed some light on the criterion for stochasticity, their results are only valid within simulated parameter range. Thus, more complete theoretical analysis is needed to have a better understanding of the mechanism of stochastic electron acceleration in the course of electron interactions with laser pulses.

In this letter, we will examine the electron dynamics in multiple laser waves by employing the Hamiltonian approach with the proper choices of canonical variables and time, such that the Hamiltonian is time independent in zero-order approximation \cite{zhang2018electron1,*zhang2018electron2}. Following \cite{mendonca1983threshold,rax1992compton,sheng2002stochastic,bochkarev2018stochastic}, we will be focused on the case where one of the laser waves is much stronger than others, which could be considered as  a perturbation. 

To simplify the expressions, in what follows we will use dimensionless variables, where $\mathbf{r}$ is normalized by the dominant laser wavenumber ($k$) and $t$ by $kc$ with $c$ being the speed of light in vacuum. The normalized parameter of laser wave, which can be described by vector potential $\mathbf{A}$, is $e\mathbf{A}/mc^2$, where $-e$ and $m$ are the electron charge and mass. In the normalized variables, we take $e=m=c=1$. 

We assume that the dominant laser wave propagates along $z$ direction and is described by the vector potential of $\mathbf{A}(v_pt-z)$, which is arbitrarily polarized in $x$ and $y$ directions (here $v_p$ is the phase velocity). For generality, we consider the perturbed laser wave propagating in the $(y,z)$ plane, determined by the vector potential of $\mathbf{A}_1(v_pt+ycos\phi+zsin\phi)$, where $\phi$ is the angle between the perturbative laser propagation direction and y-axis; and $\mathbf{A}_1$ can have three components in $x$, $y$, and $z$ directions, but $A_{1y}/A_{1z}=-tan\phi$ to ensure the orthogonality between the polarization and propagation directions. Then the electron dynamics can be described by the Hamiltonian:
\begin{equation}
\mathcal{H}\equiv\gamma=\left[1+(\mathbf{P}+\mathbf{A}+\mathbf{A_1})\right]^{1/2},\label{eqHamil-general}
\end{equation}
where $\gamma$ is the relativistic factor and $\mathbf{P}=\gamma \mathbf{v}-\mathbf{A}-\mathbf{A_1}$ is the canonical momentum. Although this Hamiltonian was widely used \cite{mendonca1983threshold,rax1992compton,LDLandaubook}, the analyses of electron dynamics accounting for both dominant and perturbative lasers were quite complicated and often incomplete.

We start our analysis with finding the proper canonical variables, Hamiltonian and effective time, such that the new Hamiltonian will be time independent when the perturbation is absent ($\mathbf{A}_1=0$). Taking into account that for $\mathbf{A}_1=0$ the combination $\gamma-v_pP_z$ is conserved, it could be considered as a candidate for the new Hamiltonian, while the phase of the dominant laser wave $\eta=v_pt-z$ can be taken as one new canonical variable. It is easy to show that for the laser field $\mathbf{A}(v_pt-z)+\mathbf{A}_1(v_pt+ycos\phi+zsin\phi)$ the canonical momentum $P_x$ is conserved so that the Hamiltonian in Eq.~(\ref{eqHamil-general}) is effectively two dimensional. Then, if we treat $(\eta,y)$ as new canonical coordinates and assume that the corresponding canonical momenta are $(\chi_\eta,\chi_y)$, while the new Hamiltonian and time are $H$ and $\tau$, the canonical transformation from the point view of least action principle \cite{landau1978course} requires that 
\begin{equation}
C(P_zdz+P_ydy-\mathcal{H} dt)=\chi_\eta d\eta+\chi_y dy-Hd\tau,\label{eqcanonical_trans}
\end{equation}
where $C$ is a constant given that the Lagrangian is not unique. The natural choice of $\tau$ is $\tau= v_pt+zsin\phi$. Substituting $\tau$ and $\eta$ into Eq.~(\ref{eqcanonical_trans}) we find $\chi_\eta=-(\gamma sin\phi+v_pP_z)$, $\chi_y=v_p(1+sin\phi)P_y$, and $H=\gamma-v_pP_z$ for $C=v_p(1+sin\phi)$. However, for convenience, we will take $\chi_\eta=+(\gamma sin\phi+v_pP_z)$, which is equivalent to choosing $\eta$ as canonical momentum while treating $\chi_\eta$ as a canonical coordinate. Then the electron dynamics can be described by $H(\chi_\eta,y,\eta,\chi_y,\tau)$, which in the new canonical variables can be found from Eqs.~(\ref{eqHamil-general}, \ref{eqcanonical_trans}). 

For the head-on colliding laser waves ($\phi=\pi/2$), $\chi_y$ is a constant. As a result, we have the following 3/2 dimensional (3/2D) Hamiltonian equations
\begin{equation}
\frac{d\chi}{d\tau}=\frac{\partial H}{\partial \eta},~\textup{and}~\frac{d\eta}{d\tau}=-\frac{\partial H}{\partial \chi},\label{eqnewHalmileq}
\end{equation}
where $\chi\equiv\chi_\eta=\gamma+v_pP_z$ and the Hamiltonian is
\begin{equation}
H(\chi,\eta,\tau)=\frac{2v_p}{v_p^2-1}\sqrt{\chi^2+(v_p^2-1)P_\perp^2}-\frac{v_p^2+1}{v_p^2-1}\chi,\label{eqHamiltonian}
\end{equation}
with $P_\perp^2=1+\sum_{i=x,y}\left[\bar{P}_i+A_i(\eta)+A_{1i}(\tau)\right]^2$ and $\bar{P}_i$ ($i=x,y$) are the conserved canonical momentum. This 3/2D Hamiltonian, which can also be obtained from the electron equations of motion, will greatly simplify our analysis in comparison with the multidimensional Hamiltonian 
\cite{mendonca1983threshold,rax1992compton} based on Eq.~(\ref{eqHamil-general}).

For simplicity and without losing the physics behind stochastic acceleration, in the following we consider the luminal case $v_p=1$, $\bar{P}_x=\bar{P}_y=0$ and two planar laser waves linearly polarized in the same direction, e.g., $\mathbf{A}=a sin(\eta)\mathbf{e}_x$ and $\mathbf{A}_1=a_1sin(k_1\tau)\mathbf{e}_x$ assuming $a_1\ll a$, where $k_1$ is the ratio of the perturbative laser frequency (or wavenumber) to that of the dominant one. Then, the Hamiltonian in Eq.~(\ref{eqHamiltonian}) degenerates to
\begin{equation}
H=\frac{1+\left[asin(\eta)+a_1sin(k_1\tau)\right]^2}{\chi}.\label{eqHamiltonian_luminal}
\end{equation}

For the unperturbed problem ($a_1=0$), the Hamiltonian in Eqs.~(\ref{eqHamiltonian}, \ref{eqHamiltonian_luminal}) is conserved and from Eqs.~(\ref{eqnewHalmileq}, \ref{eqHamiltonian_luminal}) we find the following implicit dependence $\eta(\tau)$ (we note that $\eta$ increases with $\tau$):
\begin{equation}
\tau=\frac{2+a^2}{4H^2}\left[2\eta-\frac{a^2sin(2\eta)}{2+a^2}\right]+const.,\label{eqTheta}
\end{equation}
and the frequency of unperturbed oscillation of electron canonical coordinate $\chi$:  
\begin{equation}
\omega=\frac{2\pi}{T}=\frac{4H^2}{2+a^2},\label{eqOmega}
\end{equation}
where $T=\tau(\eta=\pi)-\tau(\eta=0)$ is the period of electron oscillation.
 
From Hamiltonian in Eq.~(\ref{eqHamiltonian_luminal}) it follows that in the presence of the perturbative laser wave but for $\omega>k_1$, the electron motion is adiabatic and no electron acceleration is possible. However, when $\omega\ll k_1$ the unperturbed electron motion could resonate with the perturbative laser, $m\omega=k_1$ (where $m$ is the harmonics of unperturbed electron motion), and for the case of overlapping of the separatrices of neighbouring resonant islands, $\bar{K}=(\delta \omega+\delta \omega')/2\Delta \omega>1$, where
$\delta \omega$ and $\delta \omega '$ are their widths and $\Delta\omega$ is the distance between them, stochastic heating occurs \cite{roal1990nonlinear}. However, in what follows, we will examine the condition for an onset of stochasticity for the case $\omega/k_1\ll 1$ by using equivalent, but more convenient Chirikov-like mapping \cite{chirikov1979universal} deduced from electron equations of motion.

\begin{figure}[bt]
\centering
\begin{minipage}{1\linewidth}
\includegraphics[width=0.9\textwidth]{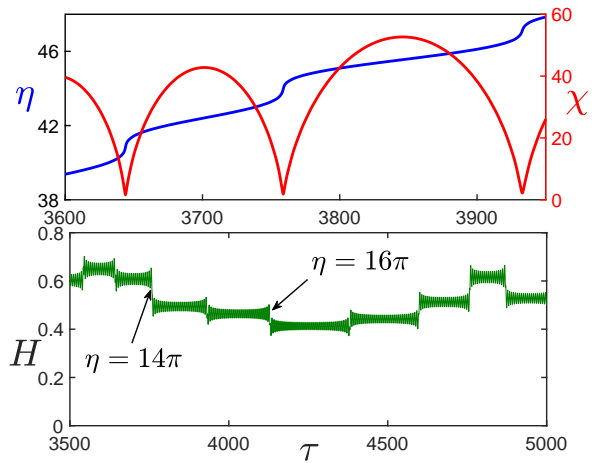}
\caption{Schematic view of electron trajectories (upper panel) and diffusion of Hamiltonian (lower panel) for $a=5$ and $a_1=0.1$.}
\label{fig-electron-trajectories}
\end{minipage}
\end{figure}

Before doing that, let us consider qualitative features of electron dynamics by exploring the Hamiltonian in Eq.~(\ref{eqHamiltonian_luminal}). Noticing that the Hamiltonian in Eq.~(\ref{eqHamiltonian_luminal}) has similar structure to that for electron in the laser and quasi-static electric fields considered in \cite{zhang2018stochastic}, we find that, for relativistic case $a> 1$, unperturbed (or weakly perturbed) electron trajectories have characteristics of zig-zag time dependence of canonical coordinate $\chi$ (e.g., see the upper panel of Fig.~\ref{fig-electron-trajectories}). This feature of electron trajectories enables a long tail of the distribution of the amplitude of $m$-harmonics, making high-$m$ island overlapping and, therefore, stochastic electron motion possible. Also, from Eq.~(\ref{eqHamiltonian_luminal}) it follows that the strongest impact, ``kicks'', on both $H$ and canonical variables by the perturbative laser occurs at a very short time near the local minimum of $\chi$ (corresponding to $\eta\approx n\pi$ with $n$ being an integer, e.g., see Fig.\ref{fig-electron-trajectories}), where the phase between electron and backward laser wave is locally minimized. Except these short periods of time $\tau$ where $\eta\approx n\pi$, the electron ``sees" only the fast phase change of the backward laser wave due to large $\chi=\gamma+P_z$ and, therefore, undergoes adiabatic oscillation.

In the following, the Chirikov-like mapping is formed by using the Hamiltonian $H_n$ and time $\tau_n$, when the electron passes through the nonadiabatic region of $\eta\approx n\pi$. Such mapping corresponds to the Poincar\'e section of electron crossing the effectively ``fixed" canonical momentum ($\eta$) plane. Let's assume that the change of the Hamiltonian due to each nonadiabatic interaction of electron with the perturbative wave (``kicks") is smaller than the Hamiltonian itself, i.e., $\Delta H=\left|H_{n+1}-H_n\right|\ll H_n$, such that the unperturbed electron trajectory $H_n(\eta,\chi)$ can be used to estimate the variation of Hamiltonian \cite{zaslavskii1987stochastic}
\begin{equation}
\Delta H_n\equiv H_{n+1}-H_n=\int_{\eta\approx n\pi}\frac{\partial H}{\partial \tau}d\tau.\label{eqDelta_H}
\end{equation}
Under the condition of $a_1\ll a$, we could keep the leading term of $\partial H/\partial \tau=2aa_1k_1sin(\eta)cos(k_1\tau)/\chi$. The fact that the main contribution to Hamiltonian variation is from $\eta\approx n\pi$ enables us to do the expansion of the integrand in Eq.~(\ref{eqDelta_H}) with respect to $\eta-n\pi$. Therefore, we have
\begin{eqnarray}
\frac{\Delta H_n}{H_n}&=&2(-1)^na_1\beta^{1/2}sin(k_1\tau_n) \label{eqDelta_H_2}\\
& &\times\int_{-\infty}^{\infty}\tilde{\eta}sin\left(\beta\tilde{\eta}+\frac{1}{3}\tilde{\eta}^3\right)d\tilde{\eta}\nonumber,
\end{eqnarray}
where $\tilde{\eta}=(\eta-n\pi)/\alpha$, $\alpha=(H_n^2/k_1a^2)^{1/3}\sim(\omega/k_1)^{1/3}\ll 1$, and $\beta=(k_1/H_n^2a)^{2/3}$. It should be noted that the fast oscillation for $\tilde{\eta}~\tilde{>}1$ justifies the extension of the integration limits to infinity. We also see that the nonadiabatic interaction of electron motion with backward laser occurs at $|\eta-n\pi|<\alpha\ll 1$ ($|\tilde{\eta}|\tilde{<}1$).

The integral in Eq.~(\ref{eqDelta_H_2}) could be expressed with the derivative of Airy function, $Ai'(\beta)$, so we have
\begin{equation}
\Delta H_n=4(-1)^{n+1}\pi a_1\beta^{1/2}Ai'(\beta)H_nsin(\psi_n),\label{eqDelta_H_3}
\end{equation}
where $\psi\equiv k_1\tau_n$. Taking into account the properties of the Airy function, it follows that the requirement of $\Delta H<H_n$ is always satisfied for $a_1\tilde{<}1$.

The time interval between two consecutive kicks is equal to the period of the unperturbed electron oscillation and thus the corresponding phase interval is determined by the Hamiltonian:
\begin{equation}
\Delta \psi_n\equiv\psi_{n+1}-\psi_n=k_1T=\frac{\pi k_1(2+a^2)}{2H_{n+1}^2}.\label{eqDelta_xi_final}
\end{equation}
As a result, rearranging Eqs.~(\ref{eqDelta_H_3}, \ref{eqDelta_xi_final}) could form symplectic mapping conserving phase volume. However, we are interested in the condition for stochasticity, which could be obtained just from Eqs.~(\ref{eqDelta_H_3}, \ref{eqDelta_xi_final}), and reads as
\begin{equation}
K=\left|\frac{d\Delta \psi_n}{dH_{n+1}}\frac{d\Delta H_n}{d\psi_n}\right|\tilde{>}1,\label{eqK_definition}
\end{equation}
where local instability leads to the mixing in phase space. If we disregard the region of phase $\psi$ where chaos appears, we arrive at  
\begin{equation}
K=4\pi^2 aa_1(2+a^2)\beta^2|Ai'(\beta)|\tilde{>}1.\label{eqK_general}
\end{equation}
A similar result can be obtained from the point of view of resonance overlapping, where one can show that $K\approx \bar{K}^2$ \cite{roal1990nonlinear}.

Introducing the function $f(\beta)=4\pi^2\beta^2|Ai'(\beta)|$, we find that $f(\beta)$ increases with $\beta$ as 
\begin{equation}
f(\beta)\approx \pi^2\beta^2,\label{eqf_smallbeta}
\end{equation}
for $\beta<1$; reaches maximum, $f_{max}\approx 8.83$, at $\beta=\beta_s\approx 1.68$; and then falls exponentially at $\beta>\beta_s$ \cite{gradshteyn2014table}:
\begin{equation}
f(\beta)\approx 2\pi^{3/2}\beta^{9/4}exp\left[-(2/3)\beta^{3/2}\right].\label{eqf_largebeta}
\end{equation}
As a result, from Eq.~(\ref{eqK_general}) we find that stochastic acceleration is only possible for $a_1>a_{s}$, where
\begin{equation}
a_{s}= \frac{f_{max}^{-1}}{a(2+a^2)}\approx\frac{0.11}{a(2+a^2)}.\label{eqThreshold-a2}
\end{equation}
We notice that the threshold in Eq.~(\ref{eqThreshold-a2}) is quite different from those in \cite{mendonca1983threshold,sheng2002stochastic}. The reason for this is that our analysis allows for finding the most stochastically ``unstable" range of $H$ (and corresponding electron kinetic energy) and, therefore, gives an exact threshold value of $a_1$ for the stochasticity onset. 

However, for $a_1$ only slightly larger than $a_s$, the stochastic acceleration occurs only within a narrow region in the vicinity of $H\approx H_s$ ($\beta\approx \beta_s$), where 
\begin{equation}
H_s\approx 0.68 \left(\frac{k_1}{a}\right)^{1/2}.\label{eqThreshold-H}
\end{equation}
For $a_1\gg a_s$ stochastic acceleration becomes possible within the range of $H$: $H_{min}<H<H_{max}$, where the lower boundary of stochasticity is due to the exponential decay of the width of resonant islands, whereas the upper one is because the distance between the neighboring resonant islands increases faster than their widths. $H_{max}$ and $H_{min}$ could be found by using asymptotic expressions (\ref{eqf_smallbeta}, \ref{eqf_largebeta}) of the function $f(\beta)$. However, we notice that the inequalities $a\gg a_1\gg a_s$ could be only satisfied for $a\gg 1$, under which we obtain:
\begin{equation}
H_{min}\approx\frac{H_s}{\sqrt{1.6+0.69ln\left(a_1/a_s\right)}}, \label{eqH_minimal}
\end{equation}
and
\begin{equation}
H_{max}\approx 1.5\left(\frac{a_1}{a_s}\right)^{3/8}H_s. \label{eqH_maxim}
\end{equation}

However, we are interested in the gain of maximum electron kinetic energy, $\gamma_{max}$, which can be expressed in the terms of $H$ as follows:
\begin{equation}
\gamma_{max}\equiv \frac{\chi+H}{2}\approx\frac{1}{2}\left(\frac{E_p}{H}+H\right), \label{eqgamma_max}
\end{equation}
 where $E_{p}=1+a^2$ is twice the ponderomotive scaling for electron energy gain in the dominant laser wave. As we see from Eq.~(\ref{eqgamma_max}), $\gamma_{max}$ can significantly exceed the ponderomotive scaling $E_p/2$ either for $H_{min}<1$ (which corresponds to the electron moving along with the dominant laser wave) or for $H_{max}>E_p$ (where the electron moves along with the perturbative laser). By using expressions (\ref{eqH_minimal}, \ref{eqH_maxim}) and neglecting numerical factors order of unity, we find that the energy of electrons moving along with dominant laser radiation exceeds the ponderomotive scaling for the case $k_1<a$ and reaches $\gamma_{max}\sim E_p(a/k_1)^{1/2}$. Whereas the energy of electrons moving along with the perturbative laser could exceed the ponderomotive scaling for the case $k_1>a^2>1$ and  $(a^2/k_1)^{4/3}<a_1/a<1$, where $\gamma_{max}\sim E_p(k_1/a^2)^{1/2}(a_1/a)^{3/8}$. We notice that $H_{min}$ and, therefore, corresponding value of $\gamma_{max}$ have a weak logarithmic dependence on the ratio $a_1/a_s>1$. Moreover, for the case $a^{-2}<k_1 < a $ and $(a^2k_1)^{-4/3} < a_1/a <1$, we have $H_{min}<1<H_{max}$ and Hamiltonian $H\sim 1$, corresponding to an initially stationary electron, is in the stochastic region. As a result, the stochastic acceleration of an electron, being initially at rest, to kinetic energy exceeding $E_p$ is possible and such energetic electron will move along with dominant laser radiation. Otherwise, initial acceleration of an electron in the direction along with (for $a_s/a<a_1/a<(k_1a^2)^{-4/3}$ and $k_1<a$) or opposite to (for $k_1>a$) the dominant laser propagation is necessary to reach the stochastic region for further acceleration.

Coming back to the expression (\ref{eqK_general}), we observe that for $\beta<\beta_s$, $K$ increases with increasing $\beta$ (which for $H_{min} <1$ corresponds to increasing electron energy). It explains the results of numerical simulations from \cite{sheng2002stochastic}, which demonstrated that the preacceleration of electrons reduces the stochastic threshold value of $a_1$ (e.g., see Fig. 3(b) in \cite{sheng2002stochastic}).
  
\begin{figure}
\centering
\subfigure{
\label{figChirikov_1}
\begin{minipage}[b]{0.23\textwidth}
\includegraphics[width=1\textwidth]{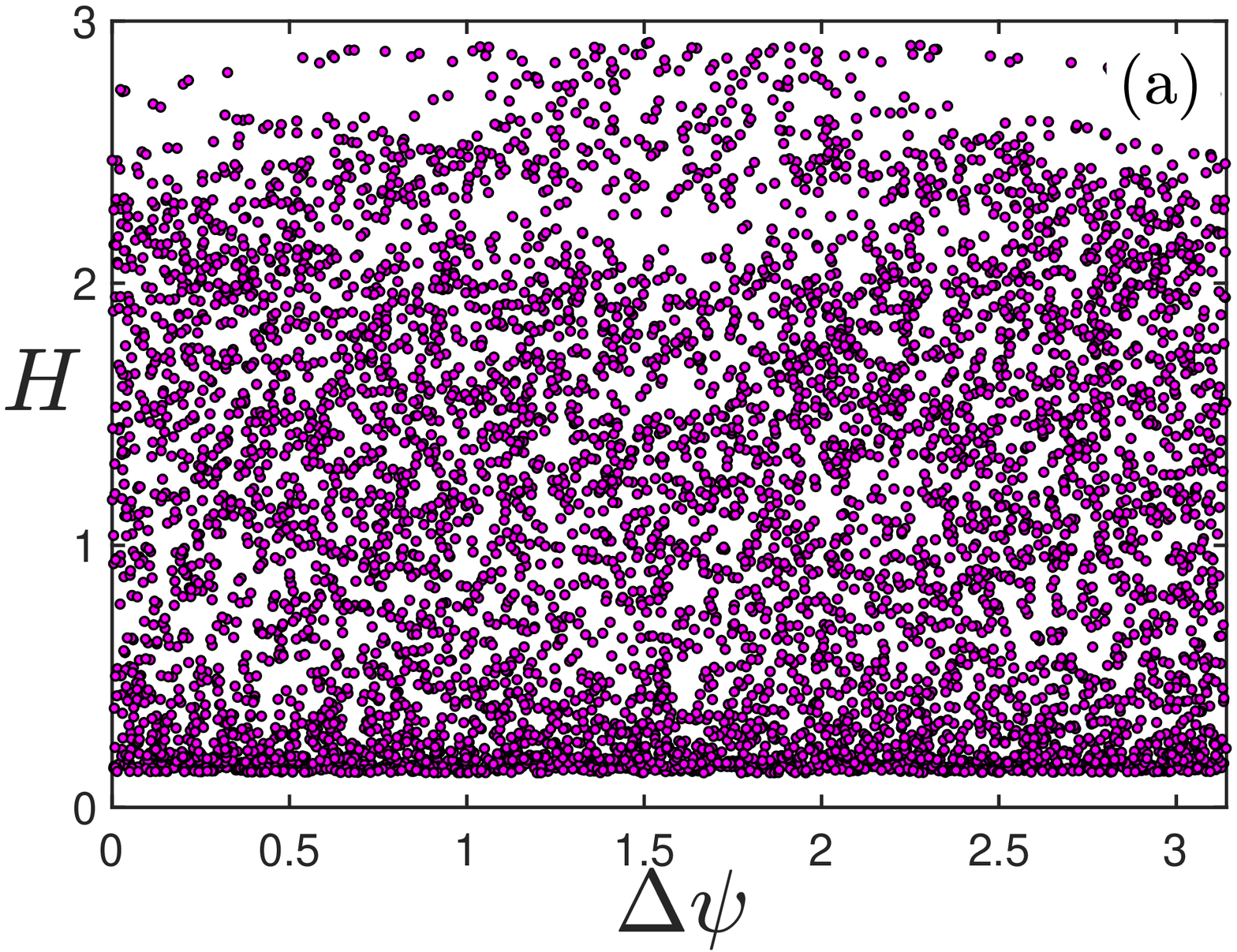}
\end{minipage}}
\subfigure{
\label{figChirikov_2}
\begin{minipage}[b]{0.23\textwidth}
\includegraphics[width=1\textwidth]{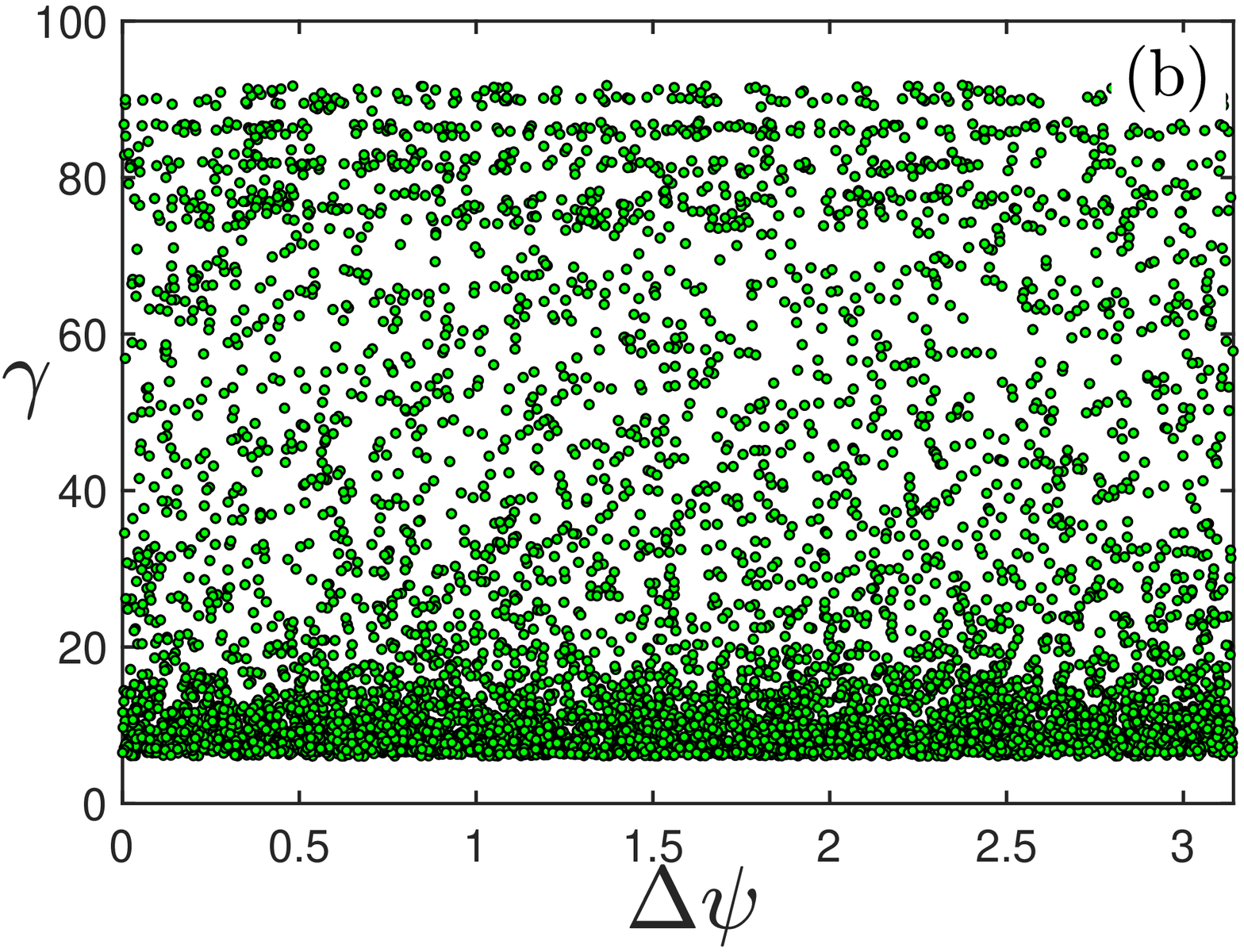}
\end{minipage}}
\caption{Poincar\'e mappings of (a) $(H,\Delta \psi)$ and (b) $(\gamma,\Delta \psi$) of electron when $\eta=n\pi+\pi/2$ for $a=5$, $a_1=0.1$, and $k_1=1$, where $\Delta \psi\equiv \psi-[\psi/\pi]\pi$.}
\label{figChirikov}
\end{figure}

\begin{figure}
\centering
\subfigure{
\label{figChirikov_3}
\begin{minipage}[b]{0.23\textwidth}
\includegraphics[width=1\textwidth]{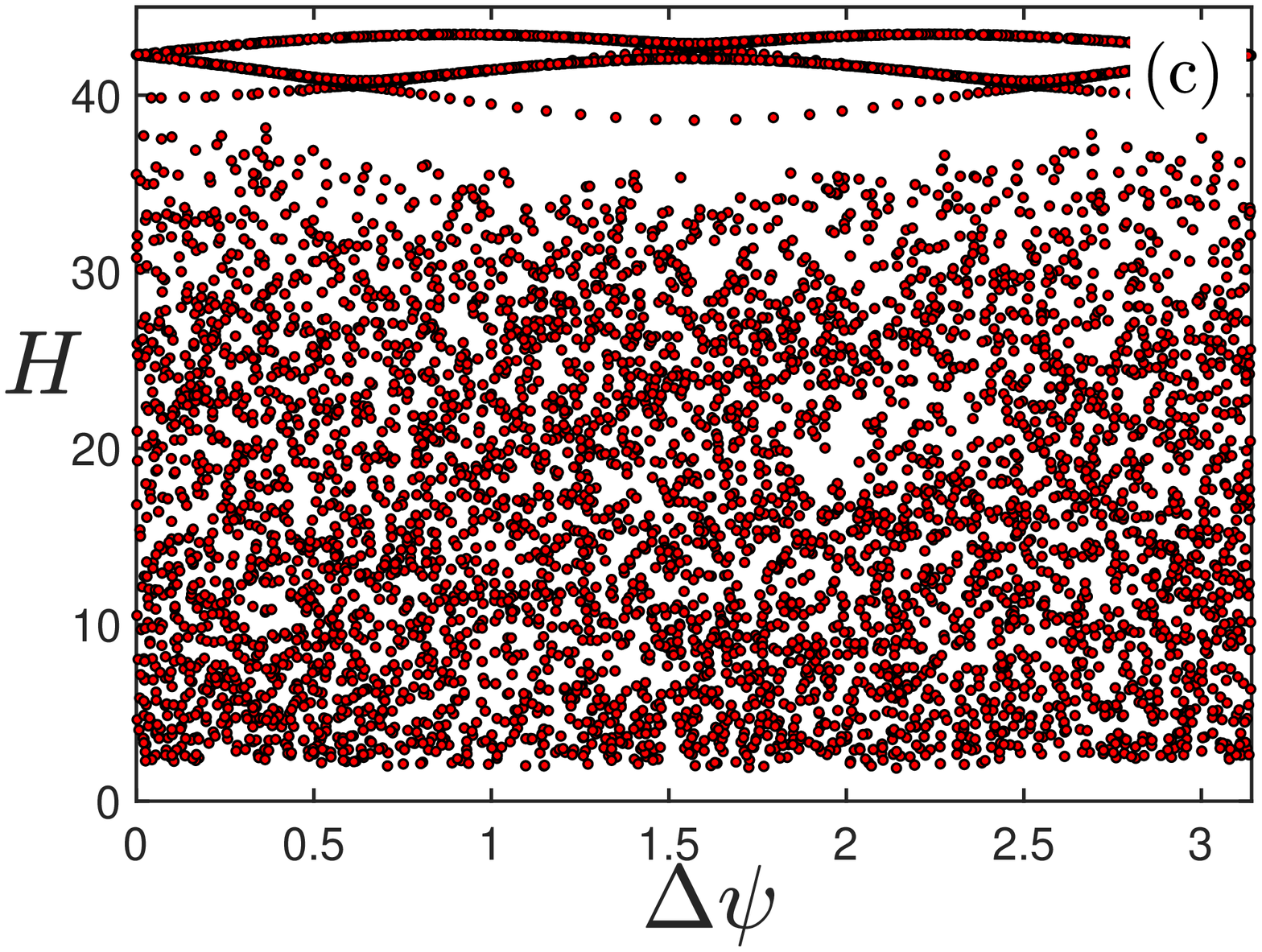}
\end{minipage}}
\subfigure{
\label{figChirikov_4}
\begin{minipage}[b]{0.23\textwidth}
\includegraphics[width=1\textwidth]{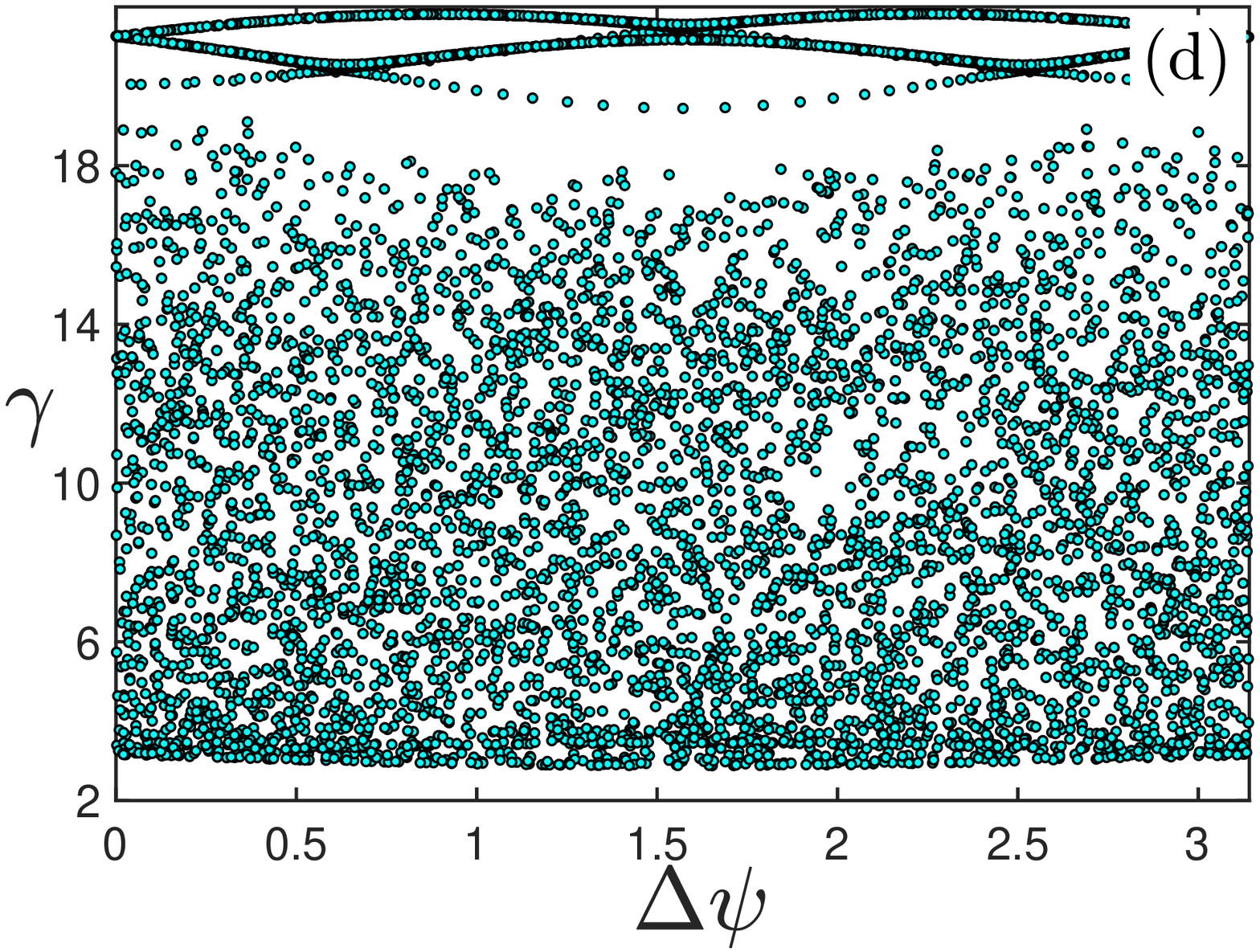}
\end{minipage}}
\caption{Poincar\'e mappings of (a) $(H,\Delta \psi)$ and (b) $(\gamma,\Delta \psi$) of electron when $\eta=n\pi+\pi/2$ for $a=3$, $a_1=0.3$, and $k_1=100$ with the same definition of $\Delta\psi$ with Fig.~\ref{figChirikov}.}
\label{figChirikov_largek1}
\end{figure}

To verify the results of our analytic consideration, we integrate Eqs.~(\ref{eqnewHalmileq}, \ref{eqHamiltonian_luminal})  numerically. The results of numerical simulations are presented in the Poincar\'e maps of ($H$, $\psi$) and the corresponding ($\gamma$, $\psi$) when $\eta=n\pi+\pi/2$. The results for $a=5$, $a_1=0.1$, and $k_1=1$ are displayed in Fig.~\ref{figChirikov}, where $k_1<a$ is satisfied and thus $\gamma_{max}\approx E_p/2H_{min}$. As one can see, a stochastic ``sea" is bounded by the KAM invariant \cite{roal1990nonlinear} at $H_{min}$ and $H_{max}$, which are, respectively, $H_{min}\approx 0.132$ and $H_{max}\approx 2.92$, and fully agree with Eqs.~(\ref{eqH_minimal}) and (\ref{eqH_maxim}). Therefore, the maximum stochastic kinetic energy ($\gamma_{max}$), which is insensitive to $a_1$ as proven in the simulations, is approximately seven times ($1/H_{min}$) larger than that without the backward wave ($E_{p}/2$). In Fig.~\ref{figChirikov_largek1} we show the results for $a=3$, $a_1=0.3$, and $k_1=100$, where $H_{min}\approx 1.87$ and $H_{max}\approx 37.5$ are, respectively, in agreement with Eq.~(\ref{eqH_minimal}) and (\ref{eqH_maxim}). For such large value of $k_1>a^2$, we see that the maximum stochastic energy satisfies $\gamma_{max}\approx H_{max}/2>E_p/2$.

In conclusion, we consider electron dynamics in the fields of colliding laser beams. We show that the proper choice of canonical variables and effective time, such that the new Hamiltonian is conserved for electrons in a dominant laser field, greatly simplifies analytical treatment of the problem. For example, for the case of counter propagating planar laser beams and dominant laser with relativistic intensity, $a > 1$, such approach allows an exhaustive analytic analysis of electron dynamics. We find that for the amplitude $a_1$ of a weaker laser ($a_1 < a$) exceeding the threshold value in Eq.~(\ref{eqThreshold-a2}), stochastic acceleration of electrons becomes possible within some range of electron kinetic energy. Maximum electron kinetic energy, which could be gained under stochastic acceleration, significantly exceeds the ponderomotive scaling for the dominant laser when the ratio, $k_1$, of perturbative to dominant laser frequencies is relatively small $k_1 < a$ (in this case, energetic electrons move in the direction of the propagation of the dominant laser beam) and for large $k_1$, such that $k_1 > a^2 > 1$, providing that $(a^2/k_1)^{4/3}<a_1/a<1$ (where energetic electrons move in the direction of the propagation of the perturbative laser beam). The results of numerical simulations, shown in Fig.~\ref{figChirikov} and Fig.~\ref{figChirikov_largek1}, are in a very good agreement with the findings from our analytic theory. We notice that the approach presented in this letter could be applied to many other cases including electron dynamics in the laser and quasi-stationary electromagnetic fields, in intense laser and Langmuir waves, etc.

This work was supported by the University of California Office of the President Lab Fee grant number LFR-17-449059.

\bibliography{main}

\begin{thebibliography}{22}%
\makeatletter
\providecommand \@ifxundefined [1]{%
 \@ifx{#1\undefined}
}%
\providecommand \@ifnum [1]{%
 \ifnum #1\expandafter \@firstoftwo
 \else \expandafter \@secondoftwo
 \fi
}%
\providecommand \@ifx [1]{%
 \ifx #1\expandafter \@firstoftwo
 \else \expandafter \@secondoftwo
 \fi
}%
\providecommand \natexlab [1]{#1}%
\providecommand \enquote  [1]{``#1''}%
\providecommand \bibnamefont  [1]{#1}%
\providecommand \bibfnamefont [1]{#1}%
\providecommand \citenamefont [1]{#1}%
\providecommand \href@noop [0]{\@secondoftwo}%
\providecommand \href [0]{\begingroup \@sanitize@url \@href}%
\providecommand \@href[1]{\@@startlink{#1}\@@href}%
\providecommand \@@href[1]{\endgroup#1\@@endlink}%
\providecommand \@sanitize@url [0]{\catcode `\\12\catcode `\$12\catcode
  `\&12\catcode `\#12\catcode `\^12\catcode `\_12\catcode `\%12\relax}%
\providecommand \@@startlink[1]{}%
\providecommand \@@endlink[0]{}%
\providecommand \url  [0]{\begingroup\@sanitize@url \@url }%
\providecommand \@url [1]{\endgroup\@href {#1}{\urlprefix }}%
\providecommand \urlprefix  [0]{URL }%
\providecommand \Eprint [0]{\href }%
\providecommand \doibase [0]{http://dx.doi.org/}%
\providecommand \selectlanguage [0]{\@gobble}%
\providecommand \bibinfo  [0]{\@secondoftwo}%
\providecommand \bibfield  [0]{\@secondoftwo}%
\providecommand \translation [1]{[#1]}%
\providecommand \BibitemOpen [0]{}%
\providecommand \bibitemStop [0]{}%
\providecommand \bibitemNoStop [0]{.\EOS\space}%
\providecommand \EOS [0]{\spacefactor3000\relax}%
\providecommand \BibitemShut  [1]{\csname bibitem#1\endcsname}%
\let\auto@bib@innerbib\@empty
\bibitem [{\citenamefont {Bauer}\ \emph {et~al.}(1995)\citenamefont {Bauer},
  \citenamefont {Mulser},\ and\ \citenamefont {Steeb}}]{bauer1995relativistic}%
  \BibitemOpen
  \bibfield  {author} {\bibinfo {author} {\bibfnamefont {D.}~\bibnamefont
  {Bauer}}, \bibinfo {author} {\bibfnamefont {P.}~\bibnamefont {Mulser}}, \
  and\ \bibinfo {author} {\bibfnamefont {W.-H.}\ \bibnamefont {Steeb}},\
  }\href@noop {} {\bibfield  {journal} {\bibinfo  {journal} {Physical review
  letters}\ }\textbf {\bibinfo {volume} {75}},\ \bibinfo {pages} {4622}
  (\bibinfo {year} {1995})}\BibitemShut {NoStop}%
\bibitem [{\citenamefont {Shvets}\ \emph {et~al.}(1998)\citenamefont {Shvets},
  \citenamefont {Fisch}, \citenamefont {Pukhov},\ and\ \citenamefont {Meyer-ter
  Vehn}}]{shvets1998superradiant}%
  \BibitemOpen
  \bibfield  {author} {\bibinfo {author} {\bibfnamefont {G.}~\bibnamefont
  {Shvets}}, \bibinfo {author} {\bibfnamefont {N.}~\bibnamefont {Fisch}},
  \bibinfo {author} {\bibfnamefont {A.}~\bibnamefont {Pukhov}}, \ and\ \bibinfo
  {author} {\bibfnamefont {J.}~\bibnamefont {Meyer-ter Vehn}},\ }\href@noop {}
  {\bibfield  {journal} {\bibinfo  {journal} {Physical review letters}\
  }\textbf {\bibinfo {volume} {81}},\ \bibinfo {pages} {4879} (\bibinfo {year}
  {1998})}\BibitemShut {NoStop}%
\bibitem [{\citenamefont {Esarey}\ \emph {et~al.}(1997)\citenamefont {Esarey},
  \citenamefont {Hubbard}, \citenamefont {Leemans}, \citenamefont {Ting},\ and\
  \citenamefont {Sprangle}}]{esarey1997electron}%
  \BibitemOpen
  \bibfield  {author} {\bibinfo {author} {\bibfnamefont {E.}~\bibnamefont
  {Esarey}}, \bibinfo {author} {\bibfnamefont {R.}~\bibnamefont {Hubbard}},
  \bibinfo {author} {\bibfnamefont {W.}~\bibnamefont {Leemans}}, \bibinfo
  {author} {\bibfnamefont {A.}~\bibnamefont {Ting}}, \ and\ \bibinfo {author}
  {\bibfnamefont {P.}~\bibnamefont {Sprangle}},\ }\href@noop {} {\bibfield
  {journal} {\bibinfo  {journal} {Physical Review Letters}\ }\textbf {\bibinfo
  {volume} {79}},\ \bibinfo {pages} {2682} (\bibinfo {year}
  {1997})}\BibitemShut {NoStop}%
\bibitem [{\citenamefont {Schroeder}\ \emph {et~al.}(1999)\citenamefont
  {Schroeder}, \citenamefont {Lee}, \citenamefont {Wurtele}, \citenamefont
  {Esarey},\ and\ \citenamefont {Leemans}}]{schroeder1999generation}%
  \BibitemOpen
  \bibfield  {author} {\bibinfo {author} {\bibfnamefont {C.}~\bibnamefont
  {Schroeder}}, \bibinfo {author} {\bibfnamefont {P.}~\bibnamefont {Lee}},
  \bibinfo {author} {\bibfnamefont {J.}~\bibnamefont {Wurtele}}, \bibinfo
  {author} {\bibfnamefont {E.}~\bibnamefont {Esarey}}, \ and\ \bibinfo {author}
  {\bibfnamefont {W.}~\bibnamefont {Leemans}},\ }\href@noop {} {\bibfield
  {journal} {\bibinfo  {journal} {Physical review E}\ }\textbf {\bibinfo
  {volume} {59}},\ \bibinfo {pages} {6037} (\bibinfo {year}
  {1999})}\BibitemShut {NoStop}%
\bibitem [{\citenamefont {Zaslavskii}\ and\ \citenamefont
  {Filonenko}(1968)}]{zaslavskii1968stochastic}%
  \BibitemOpen
  \bibfield  {author} {\bibinfo {author} {\bibfnamefont {G.}~\bibnamefont
  {Zaslavskii}}\ and\ \bibinfo {author} {\bibfnamefont {N.}~\bibnamefont
  {Filonenko}},\ }\href@noop {} {\bibfield  {journal} {\bibinfo  {journal}
  {Soviet Journal of Experimental and Theoretical Physics}\ }\textbf {\bibinfo
  {volume} {27}},\ \bibinfo {pages} {851} (\bibinfo {year} {1968})}\BibitemShut
  {NoStop}%
\bibitem [{\citenamefont {Rechester}\ and\ \citenamefont
  {Stix}(1979)}]{rechester1979stochastic}%
  \BibitemOpen
  \bibfield  {author} {\bibinfo {author} {\bibfnamefont {A.~B.}\ \bibnamefont
  {Rechester}}\ and\ \bibinfo {author} {\bibfnamefont {T.~H.}\ \bibnamefont
  {Stix}},\ }\href@noop {} {\bibfield  {journal} {\bibinfo  {journal} {Physical
  Review A}\ }\textbf {\bibinfo {volume} {19}},\ \bibinfo {pages} {1656}
  (\bibinfo {year} {1979})}\BibitemShut {NoStop}%
\bibitem [{\citenamefont {Mendonca}\ and\ \citenamefont
  {Doveil}(1982)}]{mendonca1982stochasticity}%
  \BibitemOpen
  \bibfield  {author} {\bibinfo {author} {\bibfnamefont {J.}~\bibnamefont
  {Mendonca}}\ and\ \bibinfo {author} {\bibfnamefont {F.}~\bibnamefont
  {Doveil}},\ }\href@noop {} {\bibfield  {journal} {\bibinfo  {journal}
  {Journal of Plasma Physics}\ }\textbf {\bibinfo {volume} {28}},\ \bibinfo
  {pages} {485} (\bibinfo {year} {1982})}\BibitemShut {NoStop}%
\bibitem [{\citenamefont {Mendonca}(1983)}]{mendonca1983threshold}%
  \BibitemOpen
  \bibfield  {author} {\bibinfo {author} {\bibfnamefont {J.}~\bibnamefont
  {Mendonca}},\ }\href@noop {} {\bibfield  {journal} {\bibinfo  {journal}
  {Physical Review A}\ }\textbf {\bibinfo {volume} {28}},\ \bibinfo {pages}
  {3592} (\bibinfo {year} {1983})}\BibitemShut {NoStop}%
\bibitem [{\citenamefont {Rax}(1992)}]{rax1992compton}%
  \BibitemOpen
  \bibfield  {author} {\bibinfo {author} {\bibfnamefont {J.-M.}\ \bibnamefont
  {Rax}},\ }\href@noop {} {\bibfield  {journal} {\bibinfo  {journal} {Physics
  of Fluids B}\ }\textbf {\bibinfo {volume} {4}},\ \bibinfo {pages} {3962}
  (\bibinfo {year} {1992})}\BibitemShut {NoStop}%
\bibitem [{\citenamefont {Sheng}\ \emph {et~al.}(2002)\citenamefont {Sheng},
  \citenamefont {Mima}, \citenamefont {Sentoku}, \citenamefont {Jovanovi{\'c}},
  \citenamefont {Taguchi}, \citenamefont {Zhang},\ and\ \citenamefont
  {Meyer-ter Vehn}}]{sheng2002stochastic}%
  \BibitemOpen
  \bibfield  {author} {\bibinfo {author} {\bibfnamefont {Z.-M.}\ \bibnamefont
  {Sheng}}, \bibinfo {author} {\bibfnamefont {K.}~\bibnamefont {Mima}},
  \bibinfo {author} {\bibfnamefont {Y.}~\bibnamefont {Sentoku}}, \bibinfo
  {author} {\bibfnamefont {M.}~\bibnamefont {Jovanovi{\'c}}}, \bibinfo {author}
  {\bibfnamefont {T.}~\bibnamefont {Taguchi}}, \bibinfo {author} {\bibfnamefont
  {J.}~\bibnamefont {Zhang}}, \ and\ \bibinfo {author} {\bibfnamefont
  {J.}~\bibnamefont {Meyer-ter Vehn}},\ }\href@noop {} {\bibfield  {journal}
  {\bibinfo  {journal} {Physical review letters}\ }\textbf {\bibinfo {volume}
  {88}},\ \bibinfo {pages} {055004} (\bibinfo {year} {2002})}\BibitemShut
  {NoStop}%
\bibitem [{\citenamefont {Sheng}\ \emph {et~al.}(2004)\citenamefont {Sheng},
  \citenamefont {Mima}, \citenamefont {Zhang},\ and\ \citenamefont {Meyer-ter
  Vehn}}]{sheng2004efficient}%
  \BibitemOpen
  \bibfield  {author} {\bibinfo {author} {\bibfnamefont {Z.-M.}\ \bibnamefont
  {Sheng}}, \bibinfo {author} {\bibfnamefont {K.}~\bibnamefont {Mima}},
  \bibinfo {author} {\bibfnamefont {J.}~\bibnamefont {Zhang}}, \ and\ \bibinfo
  {author} {\bibfnamefont {J.}~\bibnamefont {Meyer-ter Vehn}},\ }\href@noop {}
  {\bibfield  {journal} {\bibinfo  {journal} {Physical Review E}\ }\textbf
  {\bibinfo {volume} {69}},\ \bibinfo {pages} {016407} (\bibinfo {year}
  {2004})}\BibitemShut {NoStop}%
\bibitem [{\citenamefont {Bochkarev}\ \emph {et~al.}(2018)\citenamefont
  {Bochkarev}, \citenamefont {d'Humi{\`e}res}, \citenamefont {Tikhonchuk},
  \citenamefont {Korneev},\ and\ \citenamefont
  {Bychenkov}}]{bochkarev2018stochastic}%
  \BibitemOpen
  \bibfield  {author} {\bibinfo {author} {\bibfnamefont {S.~G.}\ \bibnamefont
  {Bochkarev}}, \bibinfo {author} {\bibfnamefont {E.}~\bibnamefont
  {d'Humi{\`e}res}}, \bibinfo {author} {\bibfnamefont {V.~T.}\ \bibnamefont
  {Tikhonchuk}}, \bibinfo {author} {\bibfnamefont {P.}~\bibnamefont {Korneev}},
  \ and\ \bibinfo {author} {\bibfnamefont {V.~Y.}\ \bibnamefont {Bychenkov}},\
  }\href@noop {} {\bibfield  {journal} {\bibinfo  {journal} {Plasma Physics and
  Controlled Fusion}\ }\textbf {\bibinfo {volume} {61}},\ \bibinfo {pages}
  {025015} (\bibinfo {year} {2018})}\BibitemShut {NoStop}%
\bibitem [{\citenamefont {Forslund}\ \emph {et~al.}(1985)\citenamefont
  {Forslund}, \citenamefont {Kindel}, \citenamefont {Mori}, \citenamefont
  {Joshi},\ and\ \citenamefont {Dawson}}]{forslund1985two}%
  \BibitemOpen
  \bibfield  {author} {\bibinfo {author} {\bibfnamefont {D.}~\bibnamefont
  {Forslund}}, \bibinfo {author} {\bibfnamefont {J.}~\bibnamefont {Kindel}},
  \bibinfo {author} {\bibfnamefont {W.}~\bibnamefont {Mori}}, \bibinfo {author}
  {\bibfnamefont {C.}~\bibnamefont {Joshi}}, \ and\ \bibinfo {author}
  {\bibfnamefont {J.}~\bibnamefont {Dawson}},\ }\href@noop {} {\bibfield
  {journal} {\bibinfo  {journal} {Physical review letters}\ }\textbf {\bibinfo
  {volume} {54}},\ \bibinfo {pages} {558} (\bibinfo {year} {1985})}\BibitemShut
  {NoStop}%
\bibitem [{\citenamefont {Landau}\ and\ \citenamefont
  {Lifshitz}(2009)}]{LDLandaubook}%
  \BibitemOpen
  \bibfield  {author} {\bibinfo {author} {\bibfnamefont {L.~D.}\ \bibnamefont
  {Landau}}\ and\ \bibinfo {author} {\bibfnamefont {E.~M.}\ \bibnamefont
  {Lifshitz}},\ }\href@noop {} {\emph {\bibinfo {title} {The Classical Theory
  of Fields, V. 2, Course of Theoretical Physics}}}\ (\bibinfo  {publisher}
  {Elsevier},\ \bibinfo {year} {2009})\BibitemShut {NoStop}%
\bibitem [{\citenamefont {Zhang}\ and\ \citenamefont
  {Krasheninnikov}(2018{\natexlab{a}})}]{zhang2018electron1}%
  \BibitemOpen
  \bibfield  {author} {\bibinfo {author} {\bibfnamefont {Y.}~\bibnamefont
  {Zhang}}\ and\ \bibinfo {author} {\bibfnamefont {S.}~\bibnamefont
  {Krasheninnikov}},\ }\href@noop {} {\bibfield  {journal} {\bibinfo  {journal}
  {Physics of Plasmas}\ }\textbf {\bibinfo {volume} {25}},\ \bibinfo {pages}
  {013120} (\bibinfo {year} {2018}{\natexlab{a}})}\BibitemShut {NoStop}%
\bibitem [{\citenamefont {Zhang}\ and\ \citenamefont
  {Krasheninnikov}(2018{\natexlab{b}})}]{zhang2018electron2}%
  \BibitemOpen
  \bibfield  {author} {\bibinfo {author} {\bibfnamefont {Y.}~\bibnamefont
  {Zhang}}\ and\ \bibinfo {author} {\bibfnamefont {S.}~\bibnamefont
  {Krasheninnikov}},\ }\href@noop {} {\bibfield  {journal} {\bibinfo  {journal}
  {Physics Letters A}\ }\textbf {\bibinfo {volume} {382}},\ \bibinfo {pages}
  {1801} (\bibinfo {year} {2018}{\natexlab{b}})}\BibitemShut {NoStop}%
\bibitem [{\citenamefont {Landau}\ and\ \citenamefont
  {Lifschic}(1978)}]{landau1978course}%
  \BibitemOpen
  \bibfield  {author} {\bibinfo {author} {\bibfnamefont {L.~D.}\ \bibnamefont
  {Landau}}\ and\ \bibinfo {author} {\bibfnamefont {E.}~\bibnamefont
  {Lifschic}},\ }\href@noop {} {\emph {\bibinfo {title} {Course of theoretical
  physics. vol. 1: Mechanics}}}\ (\bibinfo  {publisher} {Oxford},\ \bibinfo
  {year} {1978})\BibitemShut {NoStop}%
\bibitem [{\citenamefont {Sagdeev}\ \emph {et~al.}(1990)\citenamefont
  {Sagdeev}, \citenamefont {Usikov},\ and\ \citenamefont
  {Zaslavskii}}]{roal1990nonlinear}%
  \BibitemOpen
  \bibfield  {author} {\bibinfo {author} {\bibfnamefont {R.~Z.}\ \bibnamefont
  {Sagdeev}}, \bibinfo {author} {\bibfnamefont {D.}~\bibnamefont {Usikov}}, \
  and\ \bibinfo {author} {\bibfnamefont {G.~M.}\ \bibnamefont {Zaslavskii}},\
  }\href@noop {} {\emph {\bibinfo {title} {Nonlinear physics: from the pendulum
  to turbulence and chaos}}}\ (\bibinfo  {publisher} {Harwood Academic},\
  \bibinfo {year} {1990})\BibitemShut {NoStop}%
\bibitem [{\citenamefont {Chirikov}(1979)}]{chirikov1979universal}%
  \BibitemOpen
  \bibfield  {author} {\bibinfo {author} {\bibfnamefont {B.~V.}\ \bibnamefont
  {Chirikov}},\ }\href@noop {} {\bibfield  {journal} {\bibinfo  {journal}
  {Physics reports}\ }\textbf {\bibinfo {volume} {52}},\ \bibinfo {pages} {263}
  (\bibinfo {year} {1979})}\BibitemShut {NoStop}%
\bibitem [{\citenamefont {Zhang}\ \emph {et~al.}(2018)\citenamefont {Zhang},
  \citenamefont {Krasheninnikov},\ and\ \citenamefont
  {Knyazev}}]{zhang2018stochastic}%
  \BibitemOpen
  \bibfield  {author} {\bibinfo {author} {\bibfnamefont {Y.}~\bibnamefont
  {Zhang}}, \bibinfo {author} {\bibfnamefont {S.}~\bibnamefont
  {Krasheninnikov}}, \ and\ \bibinfo {author} {\bibfnamefont {A.}~\bibnamefont
  {Knyazev}},\ }\href@noop {} {\bibfield  {journal} {\bibinfo  {journal}
  {Physics of Plasmas}\ }\textbf {\bibinfo {volume} {25}},\ \bibinfo {pages}
  {123110} (\bibinfo {year} {2018})}\BibitemShut {NoStop}%
\bibitem [{\citenamefont {Zaslavskii}\ \emph {et~al.}(1987)\citenamefont
  {Zaslavskii}, \citenamefont {Natenzon}, \citenamefont {Petrovichev},
  \citenamefont {Sagdeev},\ and\ \citenamefont
  {Chernikov}}]{zaslavskii1987stochastic}%
  \BibitemOpen
  \bibfield  {author} {\bibinfo {author} {\bibfnamefont {G.}~\bibnamefont
  {Zaslavskii}}, \bibinfo {author} {\bibfnamefont {M.~Y.}\ \bibnamefont
  {Natenzon}}, \bibinfo {author} {\bibfnamefont {B.}~\bibnamefont
  {Petrovichev}}, \bibinfo {author} {\bibfnamefont {R.}~\bibnamefont
  {Sagdeev}}, \ and\ \bibinfo {author} {\bibfnamefont {A.}~\bibnamefont
  {Chernikov}},\ }\href@noop {} {\bibfield  {journal} {\bibinfo  {journal}
  {Sov. Phys. JETP}\ }\textbf {\bibinfo {volume} {66}},\ \bibinfo {pages} {496}
  (\bibinfo {year} {1987})}\BibitemShut {NoStop}%
\bibitem [{\citenamefont {Gradshteyn}\ and\ \citenamefont
  {Ryzhik}(2014)}]{gradshteyn2014table}%
  \BibitemOpen
  \bibfield  {author} {\bibinfo {author} {\bibfnamefont {I.~S.}\ \bibnamefont
  {Gradshteyn}}\ and\ \bibinfo {author} {\bibfnamefont {I.~M.}\ \bibnamefont
  {Ryzhik}},\ }\href@noop {} {\emph {\bibinfo {title} {Table of integrals,
  series, and products}}}\ (\bibinfo  {publisher} {Academic press},\ \bibinfo
  {year} {2014})\BibitemShut {NoStop}%
\end{thebibliography}%

\end{document}